\documentclass[fleqn,usenatbib]{mnras}

\usepackage{newtxtext,newtxmath}
\usepackage{subcaption}
\usepackage{placeins}
\usepackage{float}
\usepackage{bm}
\usepackage[T1]{fontenc}
\usepackage{ae,aecompl}
\usepackage{enumitem}

\usepackage{graphicx}	
\usepackage{amsmath}	
\usepackage{longtable}
\usepackage{rotating}
\usepackage{amsmath}
\usepackage{algorithm}
\usepackage{algpseudocode}
\usepackage{tikz}
\usepackage{lipsum} 

\title[MCMC spectral parameters of G339.884‑1.259]{Bayesian estimation of spectral parameters of the 6.7‑GHz methanol maser G339.884‑1.259 from GRAO observations}

\author[T. Ansah-Narh et al.]{Theophilus Ansah-Narh$^{1}$\thanks{E-mail: philusnarh@gmail.com (TAN)},
Stephen Sottie$^{2}$,
Nia Imara$^{3}$,
Emmanuel Proven-Adzri$^{1}$
\\
$^{1}$Ghana Space Science \& technology Institute, Ghana Atomic Energy Commission, P. O. Box LG 80, Legon-Accra, Ghana.\\
$^{2}$ Department of Physics and Electronics, Rhodes University, P. O.  Box 94, Grahamstown 6140, South Africa\\
$^{3}$ Astronomy and Astrophysics Department, University of California Santa Cruz, Santa Cruz, CA, USA.
}

\date{Accepted XXX. Received YYY; in original form ZZZ}

\pubyear{2015}

\begin{document}
\label{firstpage}
\pagerange{\pageref{firstpage}--\pageref{lastpage}}
\maketitle

\begin{abstract}
Accurate decomposition of methanol maser spectra is essential for constraining the kinematics and physical conditions of high-mass star-forming regions, particularly in complex blended spectra where small differences in component structure can alter physical interpretation. Conventional profile fitting approaches often rely on fixed Gaussian decompositions that do not fully capture non-Gaussian spectral structure and provide limited statistical characterisation of parameter uncertainties. To address this, we develop a Bayesian spectral decomposition framework that models the 6.7\,GHz methanol maser emission using alternative Gaussian, Lorentzian, and Voigt profile families, with parameter posteriors inferred through Markov Chain Monte Carlo sampling. This probabilistic framework enables simultaneous model comparison, uncertainty quantification, and statistically consistent estimation of spectral components.
Application to the methanol maser G339.884$-$1.259 observed with the Ghana Radio Astronomy Observatory reveals a complex multi-component spectrum composed of seven velocity-coherent features with tightly constrained parameters. Comparative analysis demonstrates that models incorporating both Doppler-like cores and Lorentzian wing structure provide the statistically preferred representation of the observed spectra, yielding the lowest information criteria (AIC $\approx 1.98 \times 10^{4}$; BIC $\approx 1.99 \times 10^{4}$), the smallest residual errors (RMSE $\approx 11.1$\,Jy), and the highest goodness-of-fit ($R^{2} \approx 0.985$). In contrast, purely Gaussian or Lorentzian representations leave systematic residual structure near the line wings and strongest maser components. Elevated reduced $\chi^{2}_{\nu}$ values across all tested models further indicate that unresolved spectral substructure, non-ideal noise properties, and intrinsic line-profile complexity remain important limitations in single-dish maser decomposition.
These results demonstrate that Bayesian inference provides a robust and reproducible framework for analysing complex maser spectra while simultaneously quantifying parameter uncertainties and model limitations. The methodology is readily extendable to other molecular line studies and establishes a pathway toward integrating statistically rigorous spectral modelling with high-resolution interferometric observations to better constrain the dynamics and environments of massive star formation.
\end{abstract}




\section{Introduction}
\label{paper:intro}

The 6.7\,GHz transition of methanol (CH$_3$OH) produces some of the brightest maser lines in the Galaxy, with flux densities often reaching hundreds to thousands of Janskys. Since its discovery by \citet{1991ApJ...380L..75M}, this transition has become a fundamental tool for identifying and characterising regions of high-mass star formation. A growing body of evidence indicates that Class\,II methanol masers, of which the 6.7\,GHz line is the dominant representative, are exclusively associated with the early stages of massive star formation, typically tracing phases preceding the development of ultracompact H\,\textsc{ii} regions \citep{2010MNRAS.401.2219B, 2010MNRAS.409..913G, 1996PhDT........15E}. Their high brightness and compact emission make them ideal targets for high-resolution studies of the kinematics and physical conditions in the innermost environments of massive protostars, including accretion disks, jets, and outflows.

The source G339.884$-$1.259 stands out as one of the most prominent and well-studied 6.7\,GHz methanol masers in the southern sky. It is a bright and complex source, with interferometric observations revealing a linear distribution of maser spots and a monotonic velocity gradient that initially suggested the presence of a rotating circumstellar disk around a massive star \citep{1996MNRAS.279..101E}. 
Later mid-infrared imaging, however, resolved the emission into three distinct sources, challenging the single-disk interpretation and pointing instead to a more complicated environment involving a collimated jet or outflow \citep{2002ApJ...564..327D}. The region also hosts the brightest known H$_2$CO maser in the Milky Way and displays a rich spectrum at millimetre wavelengths \citep{chen2017brightest}. This source thus offers a valuable laboratory for studying the complex interplay between accretion, outflow, and maser excitation in massive star formation.

Despite the wealth of information that can be extracted from the 6.7\,GHz methanol maser line, the standard approach to analysing its spectral profile has remained largely unchanged for three decades. Investigators typically inspect the spectrum manually, identify a set of visually distinct peaks, and then fit a sum of Gaussian functions to the data, often using least-squares algorithms. While widely used and often effective, such approaches can face important limitations when applied to complex blended maser spectra. First, the choice of the number of components and their initial parameters is subjective; different researchers may obtain different decompositions from the same spectrum, and small changes in component selection can substantially alter the fitted parameters.
Second, the Gaussian model itself is only an approximation. In maser environments, deviations from purely Gaussian profiles may arise from several mechanisms, including unresolved velocity gradients, turbulence, partial saturation, radiative transfer effects, spectral blending of nearby maser spots, and potentially damping-related processes \citep{elitzur1992astronomical}. Consequently, a purely Gaussian representation may not always capture extended line wings or subtle asymmetries present in high signal-to-noise spectra. Restricting the decomposition to Gaussian components alone may therefore overlook non-Gaussian structure in the observed profiles and bias the inferred component parameters.
Third, conventional fitting methods, whether based on Gaussian or Voigt functions, may in some cases return primarily point estimates
of the parameters without fully characterising the associated uncertainties or parameter correlations. In complex blended spectra,
limited uncertainty quantification can make it more difficult to assess whether a difference in, for example, central velocity between two components is physically meaningful or influenced by the fitting procedure.

A flexible statistical framework for addressing these issues is
provided by Bayesian inference implemented through Markov Chain Monte Carlo (MCMC) sampling. In this framework, the model
parameters are treated as random variables, and the goal is to obtain their posterior probability distributions given the data and
any prior information. MCMC methods have gained considerable traction in astronomy for tasks ranging from exoplanet orbit fitting
to stellar population synthesis
\citep{2025A&A...704A.323P,2023ApJ...959...20J,
2017AAS...22914602B, 2012ApJ...745..198H,
acquaviva2011sed, 2005AJ....129.1706F}, but their application to maser spectral decomposition remains rare. A Bayesian approach
offers several advantages: it enables direct estimation of parameter uncertainties in the form of posterior credible intervals, allows
model complexity to be assessed via information criteria such as the Akaike Information Criterion (AIC) or the Bayesian Information Criterion (BIC), and permits the incorporation of physically motivated priors (e.g., that line widths must be positive). In the specific context of maser spectroscopy, Bayesian MCMC provides a
statistically principled way to decompose a complex, blended
spectrum into its constituent velocity components while
simultaneously characterising the precision and covariance of the derived parameters.

In this paper, we apply a Bayesian MCMC method to the 6.7\,GHz methanol maser spectrum of G339.884$-$1.259, using data obtained with the 32-m radio telescope of the Ghana Radio Astronomy Observatory (GRAO). The observations were conducted in a standard position-switching mode, and the spectrum was extracted after subtraction of an off-source (blank sky) measurement. We model the spectrum as a sum of Voigt profiles, each described by an amplitude, a central velocity, a Gaussian width ($\sigma$) associated with turbulent broadening, and a Lorentzian width ($\gamma$) associated with damping. The number of components is determined via automated peak detection, but the final fit allows the parameters of all components to be adjusted simultaneously within the Bayesian framework. The least-squares solution from a preliminary fit is used as the starting point for the MCMC sampling, which we run with 100 walkers and 2000 steps, discarding an initial burn-in phase and thinning the chains to reduce correlation. The posterior distributions provide median values and 95\% credible intervals for the amplitude, central velocity, and full width at half maximum (FWHM) of each detected component.

Our primary scientific aim is to obtain a statistically rigorous set of spectral parameters for the maser components of G339.88$-$1.26, thereby providing a solid basis for kinematic and physical interpretations. A secondary aim is to demonstrate the feasibility and advantages of Bayesian MCMC decomposition for maser spectroscopy, using a modest-sized single-dish telescope. The method is general and can be applied to any maser or molecular line spectrum, offering a pathway to more reproducible and better-characterised measurements. The structure of the paper is as follows. Section~\ref{sec:obs} describes the observations and data reduction. Section~\ref{sec:methodology} presents the Bayesian MCMC methodology, including the composite Voigt model, the peak detection algorithm, and the fitting procedure. Section~\ref{sec:results} reports the results: the fitted components, their parameters, and a comparison between the Voigt and Gaussian models. Section~\ref{sec:discuss} discusses the implications of the results for the kinematics and physical conditions in G339.88$-$1.26, as well as the broader methodological lessons. Section~\ref{sec:conc} summarises the conclusions.

\section{Observations and Data Reduction} \label{sec:obs}

\subsection{GRAO 32-m Telescope}

The observations were carried out on 2021 June 9 using the 32-m radio telescope at the Ghana Radio Astronomy Observatory (GRAO), located at latitude $5^\circ45'1.3''$ N and longitude $0^\circ18'18.5''$ W. The telescope employs a beam waveguide optical system with a primary reflector diameter of 32 m and an aperture efficiency of 77 per cent at 6.7 GHz (GRAO Technical Fact Sheet). The receiver is a single-channel, ambient-temperature system with a noise temperature of approximately 125 K at 5 GHz, though the value at 6.7 GHz is expected to be similar. The half-power beamwidth at this frequency is $0.09^\circ$. The observations used a standard position-switching mode, alternating between the target source (ON) and a nearby blank sky position (OFF) to remove atmospheric and receiver contributions. The total integration time was approximately 20 minutes for the ON scan and a similar duration for the OFF scan. The backend spectrometer was configured with a bandwidth of 2 MHz, centred on the rest frequency of the 6.7 GHz methanol maser (6668.518 MHz). The number of spectral channels was 4096, yielding a channel spacing of 488 Hz, which corresponds to a velocity resolution of approximately 0.022 km s$^{-1}$ at this frequency.

\subsection{Data Reduction}

The raw data were stored in HDF5 format, with the backend spectral data located in the \texttt{Data/VisData} dataset. The initial extraction and preprocessing were performed using locally developed \texttt{Python} routines based on standard scientific libraries. Although the backend stored complex-valued samples internally, the present analysis used the corresponding total-power spectra derived from the channel power amplitudes. The data contained multiple short integrations; these were averaged along the time axis to produce mean spectra for both the ON and OFF scans. No full polarimetric calibration was performed. Significant linear polarisation has previously been reported for G339.884$-$1.259 \citep{2008A&A...480..767D}, and methanol masers may also exhibit circular polarisation associated with Zeeman splitting \citep{2008A&A...484..773V}. However, the present analysis focuses on the total intensity spectrum (Stokes~I), which is sufficient for the spectral decomposition and line-profile analysis undertaken here.

The OFF source spectrum was subtracted from the ON source spectrum channel by channel to remove residual baseline variations and any instrumental offsets. The resulting difference spectrum represents the maser line emission. The observations were obtained at a fixed observing frequency centred on the 6.7\,GHz methanol transition. To convert the observed frequencies to a kinematic velocity scale, we used the radio definition
\begin{equation}
v = c \frac{f_0 - f}{f_0},
\label{eq:radio_velocity}
\end{equation}
where $c$ is the speed of light in km s$^{-1}$, $f_0 = 6668.518$ MHz is the rest frequency, and $f$ is the observed frequency of each channel. The spectrum was then transformed to the Local Standard of Rest kinematic (LSRK) frame using the observing coordinates and observation time. Because the observations were obtained during engineering commissioning of the telescope backend, small residual uncertainties in the local oscillator and frequency calibration could not be independently characterised. To minimise possible systematic offsets, the velocity axis was therefore refined by aligning the strongest maser feature with its well-established literature velocity near $-35$ km s$^{-1}$ \citep{2011MNRAS.417.1964C, 1993ApJ...412..222N}. Consequently, while the relative velocity separations between spectral components are robust, the present data are not suitable for measuring absolute secular velocity drifts at the level of a few hundred m\,s$^{-1}$. The final calibrated spectrum spans an LSRK velocity range from $-45$ to $-20$ km s$^{-1}$, with maser emission concentrated between $-35$ and $-22$ km s$^{-1}$.
Further baseline fitting (for example, a polynomial) was not required, as the OFF-subtracted spectrum exhibited a flat continuum within the noise level.

The absolute flux density scale was derived from the telescope’s system temperature and gain, rather than from an independent calibrator observation. At the time of these observations, the GRAO 32-m telescope was in Phase~II engineering commissioning; a standard flux calibrator (for instance, a noise diode or a known astronomical point source) was not yet routinely observed. Consequently, we did not have a contemporaneous measurement of the telescope’s response on an absolute scale.
We therefore used the theoretical gain of the telescope at 6.7 GHz, computed from its aperture efficiency. The effective area $A_{\mathrm{eff}}$ is given by
\begin{equation}
A_{\mathrm{eff}} = \eta \, \pi (D/2)^2,
\label{eq:effective_area}
\end{equation}
where $\eta = 0.77$ (aperture efficiency) and $D = 32$ m (diameter). This yields $A_{\mathrm{eff}} \approx 619.3$ m$^2$. The gain in units of K Jy$^{-1}$ is then
\begin{equation}
G = \frac{A_{\mathrm{eff}}}{2k} \times 10^{-26},
\label{eq:gain}
\end{equation}
where $k = 1.380649 \times 10^{-23}$ J K$^{-1}$. The result is $G \approx 0.224$ K Jy$^{-1}$; that is, 1 Jy of source flux density produces a temperature increase of 0.224 K, or equivalently 1 K corresponds to approximately $4.46$ Jy.
The system temperature during the observation was approximately $T_{\mathrm{sys}} = 125$ K (based on the GRAO specification sheet, typical for the 5--7 GHz band). The raw power measured from the ON--OFF subtraction is proportional to the source antenna temperature $T_{\mathrm{A}}$. Using the gain factor, we converted the power to flux density as
\begin{equation}
S_{\mathrm{Jy}} = \frac{T_{\mathrm{A}}}{G}
\label{eq:flux_conversion}
\end{equation}

As an independent consistency check, we compared the peak flux of the brightest maser component obtained from this conversion with the value of 1400 Jy reported by \citet{1996MNRAS.279..101E}. The two agreed to within approximately 15 per cent. Given the uncertainty in the exact system temperature and the absence of a primary calibrator, we consider this agreement satisfactory and indicative that our derived flux scale is reasonable for the purpose of line profile analysis. However, the absolute flux densities should be treated with caution, and we emphasise that the primary results of this paper (central velocities, line widths, and component shapes) are unaffected by this scaling.
Finally, the noise level, estimated from line-free channels, was approximately 15 Jy rms.

\section{Methodology} \label{sec:methodology}

\subsection{Composite line profile models}

The observed spectral profile of a methanol maser reflects the combined influence of Doppler motions, radiative transfer effects, spectral blending, velocity gradients, turbulence, saturation behaviour, and potentially additional non-Gaussian broadening processes.
Each component can be described by a functional form. To capture the complexity of the maser emission, we model the spectrum as a sum of individual line profiles, each characterised by a central velocity, an amplitude, and a width parameter. We consider three families: Gaussian, Lorentzian, and Voigt.

The Gaussian profile is appropriate when the dominant line broadening arises from thermal motions and microturbulence. In many maser environments, Doppler motions associated with thermal and microturbulent processes can produce approximately Gaussian line cores, particularly in unsaturated or weakly saturated regimes \citep{Gray2012}. For a single component, the Gaussian profile as a function of velocity \(v\) is
\begin{equation}
    G(v; A, \mu, \sigma) = \frac{A}{\sigma \sqrt{2\pi}} \exp\left[-\frac{(v-\mu)^2}{2\sigma^2}\right], \label{eq:gauss}
\end{equation}
where \(A\) is the integrated area (in Jy km s\(^{-1}\)), \(\mu\) the central LSR velocity (km s\(^{-1}\)), and \(\sigma\) the Gaussian standard deviation (km s\(^{-1}\)). The full width at half maximum (FWHM) is
\begin{equation}
    \Delta v_{\rm G} = 2\sqrt{2\ln 2}\,\sigma \approx 2.35482\,\sigma. \label{eq:fwhm_g}
\end{equation}

Lorentzian functions provide a convenient representation of extended spectral wings and non-Gaussian structure. In astrophysical masers, such behaviour may arise from several mechanisms, including radiative transfer effects, unresolved kinematic blending, saturation-induced re-broadening, velocity gradients, turbulence, or damping-related processes. The relative importance of collisional and radiative effects depends strongly on the local physical conditions and pumping environment of the maser \citep{Gray2012, 2005MNRAS.360..533C,elitzur1992astronomical}. The Lorentzian term used here is therefore interpreted phenomenologically rather than as a direct measurement of classical pressure broadening. The Lorentzian profile is defined as;
\begin{equation}
    L(v; A, \mu, \gamma) = \frac{A}{\pi\gamma} \, \frac{1}{1 + \left(\frac{v-\mu}{\gamma}\right)^2}, \label{eq:lorentz}
\end{equation}
where \(\gamma\) is the half width at half maximum (HWHM). The FWHM is then
\begin{equation}
    \Delta v_{\rm L} = 2\gamma. \label{eq:fwhm_l}
\end{equation}

In many real astrophysical environments, both Doppler and non-Gaussian broadening mechanisms can contribute simultaneously to the observed spectral profile. The Voigt profile, which is the convolution of a Gaussian and a Lorentzian, provides a flexible representation of spectral profiles exhibiting both narrow cores and extended wings and is widely used in spectroscopic modelling \citep{Gray2012}. It is particularly useful for describing spectral profiles that exhibit both narrow cores and extended non-Gaussian wings, irrespective of the specific underlying physical origin of those deviations. The Voigt profile is given by the convolution integral
\begin{equation}
    V(v; A, \mu, \sigma, \gamma) = \int_{-\infty}^{\infty} G(v' - v; \sigma) \, L(v - v'; \gamma) \, \mathrm{d}v'. \label{eq:voigt_conv}
\end{equation}
This integral has no closed form but can be expressed using the Faddeeva function \(w(z)\):
\begin{equation}
    V(v; A, \mu, \sigma, \gamma) = \frac{A}{\sigma \sqrt{2\pi}} \, \Re\left[w(z)\right], \qquad 
    z = \frac{v - \mu + \mathrm{i}\gamma}{\sigma \sqrt{2}}, \label{eq:voigt_faddeeva}
\end{equation}
where $\Re[w(z)]$ denotes the real part. For the FWHM of a Voigt component we adopt the approximation of \cite{1977JQSRT..17..233O}:
\begin{equation}
    \Delta v_{\rm V} \approx \frac{\Delta v_{\rm L}}{2} + \sqrt{\frac{\Delta v_{\rm L}^2}{4} + \Delta v_{\rm G}^2}. \label{eq:fwhm_v}
\end{equation}

In the present work, the Voigt profile is adopted as a phenomenological model for spectral decomposition and should not be interpreted as uniquely identifying specific microscopic broadening mechanisms.
Within the Voigt formalism, the Gaussian and Lorentzian terms provide a flexible parametrisation of core and wing behaviour in the observed spectrum. These components should not necessarily be interpreted as uniquely tracing isolated physical broadening mechanisms. The Voigt model therefore provides a more flexible representation of complex spectral morphology than purely Gaussian or purely Lorentzian forms. However, to assess the empirical importance of non-Gaussian wing structure, we also fit composite pure Gaussian and pure Lorentzian models, i.e., the spectrum is modelled as a sum of functions of the type defined in Eq.~\eqref{eq:gauss} or Eq.~\eqref{eq:lorentz} alone. This comparison allows us to determine whether the additional complexity of the Voigt model is justified by the data.

\subsection{Bayesian inference framework}

To quantify the uncertainties inherent in decomposing the maser spectrum into individual velocity components, we adopt a Bayesian approach. This framework treats the model parameters \(\Theta\) (amplitudes, central velocities, and widths for all components) as random variables whose posterior probability distribution, given the observed spectrum \(D = \{(v_i, y_i)\}_{i=1}^{N}\), is obtained via Bayes' theorem:
\begin{equation}
    P(\Theta \mid D, M) = \frac{P(D \mid \Theta, M) \, P(\Theta \mid M)}{P(D \mid M)}, \label{eq:bayes}
\end{equation}
where \(M\) denotes the chosen model (e.g., a sum of \(K\) Voigt profiles), \(P(D \mid \Theta, M)\) is the likelihood, \(P(\Theta \mid M)\) the prior, and \(P(D \mid M)\) the evidence. The observed data consist of velocity channels \(v_i\) and corresponding flux densities \(y_i\). The evidence acts as a normalisation constant that does not affect parameter estimation; it is therefore ignored when sampling the posterior.

The likelihood quantifies how well the model reproduces the observed data. We assume that the residuals between the data and the model are independent and identically distributed according to a Gaussian with zero mean and constant standard deviation \(\varepsilon\). The value of \(\varepsilon\) is estimated from the RMS noise measured in line-free channels of the subtracted spectrum. The likelihood is then
\begin{equation}
    P(D \mid \Theta, M) = \prod_{i=1}^{N} \frac{1}{\sqrt{2\pi\varepsilon^2}} \exp\left[-\frac{\left(y_i - f_M(v_i; \Theta)\right)^2}{2\varepsilon^2}\right], \label{eq:likelihood}
\end{equation}
with \(f_M(v; \Theta)\) the model prediction (sum of component profiles). The equivalent log-likelihood is
\begin{equation}
    \ln P(D \mid \Theta, M) = -\frac{N}{2}\ln(2\pi\varepsilon^2) - \frac{1}{2\varepsilon^2} \sum_{i=1}^{N} \left(y_i - f_M(v_i; \Theta)\right)^2. \label{eq:loglike}
\end{equation}

Prior knowledge about the parameters is expressed through the prior distribution \(P(\Theta \mid M)\). We adopt weakly informative priors that are broad enough to let the data dominate the posterior yet sufficiently constrained to exclude physically impossible values. For the amplitude of any component we use a uniform prior bounded below by zero and above by a large cutoff (\(10^4\)~Jy~km~s\(^{-1}\)) that does not affect the posterior. The centre of each component is assigned a uniform prior over an interval of \(\pm5.0\)~km~s\(^{-1}\) around its initial guess from peak detection, which comfortably encompasses the entire maser emission range observed in G339.884-1.259 (approximately \(-35\) to \(-22\)~km~s\(^{-1}\)). For width parameters, we apply a uniform prior on \([0.01,\,2.00]\)~km~s\(^{-1}\) to both the Gaussian standard deviation \(\sigma\) (in the Gaussian and Voigt models) and the Lorentzian half-width \(\gamma\) (in the Lorentzian and Voigt models). These bounds exclude unphysically narrow or excessively broad lines while remaining uninformative over the expected range for methanol masers, where typical FWHM values lie between \(0.1\) and \(1.5\)~km~s\(^{-1}\). For a given model, only the parameters that actually appear (e.g., for the Gaussian model: \(A\), \(\mu\), \(\sigma\); for the Lorentzian model: \(A\), \(\mu\), \(\gamma\); for the Voigt model: \(A\), \(\mu\), \(\sigma\), \(\gamma\)) are assigned priors; the irrelevant parameters are simply omitted. All priors are independent, so the joint prior is the product of the individual priors:
\begin{equation}
    P(\Theta \mid M) = \prod_{j} P(A_j) P(\mu_j) P(\sigma_j) P(\gamma_j). \label{eq:prior}
\end{equation}
For models lacking a particular parameter (e.g., \(\gamma\) in the pure Gaussian model), the corresponding factor is omitted from the product.

Substituting the likelihood Eq.~\eqref{eq:likelihood} and the prior Eq.~\eqref{eq:prior} into Bayes' theorem Eq.~\eqref{eq:bayes} yields the unnormalised posterior distribution:
\begin{equation}
    P(\Theta \mid D, M) \propto \exp\left[-\frac{1}{2\varepsilon^2} \sum_{i=1}^{N} \left(y_i - f_M(v_i; \Theta)\right)^2\right] \; \prod_{\theta \in \Theta} P(\theta). \label{eq:posterior}
\end{equation}
This distribution is high-dimensional and does not have a closed analytical form. We therefore sample it numerically using MCMC, as described next.

\subsection{Markov Chain Monte Carlo: Metropolis-Hastings algorithm}

MCMC generates a sequence of samples whose stationary distribution
is the target posterior. We employ the Metropolis-Hastings
algorithm, a general method for constructing such Markov chains.
In this framework, the posterior distribution is sampled by
iteratively proposing candidate parameter sets and accepting or
rejecting them according to their relative posterior probabilities.

Define the target distribution
\(\pi(\Theta) = P(\Theta \mid D, M)\). Let
\(q(\Theta' \mid \Theta)\) denote the proposal distribution used
to generate a candidate state \(\Theta'\) from the current state
\(\Theta\). The algorithm proceeds as follows:

\begin{enumerate}
    \item Initialize \(\Theta^{(0)}\) (e.g., from the least‑squares fit).
    \item For \(t = 1, 2, \dots, T\):
        \begin{enumerate}
            \item Propose \(\Theta' \sim q(\Theta' \mid \Theta^{(t-1)})\).
            \item Compute the acceptance probability
            \begin{equation}
                \alpha = \min\left(1,\; \frac{\pi(\Theta') \, q(\Theta^{(t-1)} \mid \Theta')}{\pi(\Theta^{(t-1)}) \, q(\Theta' \mid \Theta^{(t-1)})}\right). \label{eq:acceptance}
            \end{equation}
            \item Draw \(u \sim \mathrm{Uniform}(0,1)\).
            \item If \(u < \alpha\), set \(\Theta^{(t)} = \Theta'\) (accept); otherwise set \(\Theta^{(t)} = \Theta^{(t-1)}\) (reject).
        \end{enumerate}
\end{enumerate}

For our implementation we use a random‑walk Metropolis sampler with a multivariate Gaussian proposal:
\begin{equation}
    q(\Theta' \mid \Theta) = \mathcal{N}(\Theta' \mid \Theta, \Sigma), \label{eq:proposal}
\end{equation}
where \(\Sigma\) is a diagonal covariance matrix. The proposal is symmetric, so \(q(\Theta' \mid \Theta) = q(\Theta \mid \Theta')\) and Eq.~\eqref{eq:acceptance} simplifies to the Metropolis ratio:
\begin{equation}
    \alpha = \min\left(1,\; \frac{\pi(\Theta')}{\pi(\Theta^{(t-1)})}\right). \label{eq:metropolis_ratio}
\end{equation}
The step sizes (diagonal elements of \(\Sigma\)) are tuned during a short pilot run to achieve an acceptance rate of approximately $0.25$, which is optimal for high‑dimensional problems \citep{chepelianskii2023metropolis, gelman1996efficient}

\begin{figure*}
\centering
\includegraphics[width=\textwidth]{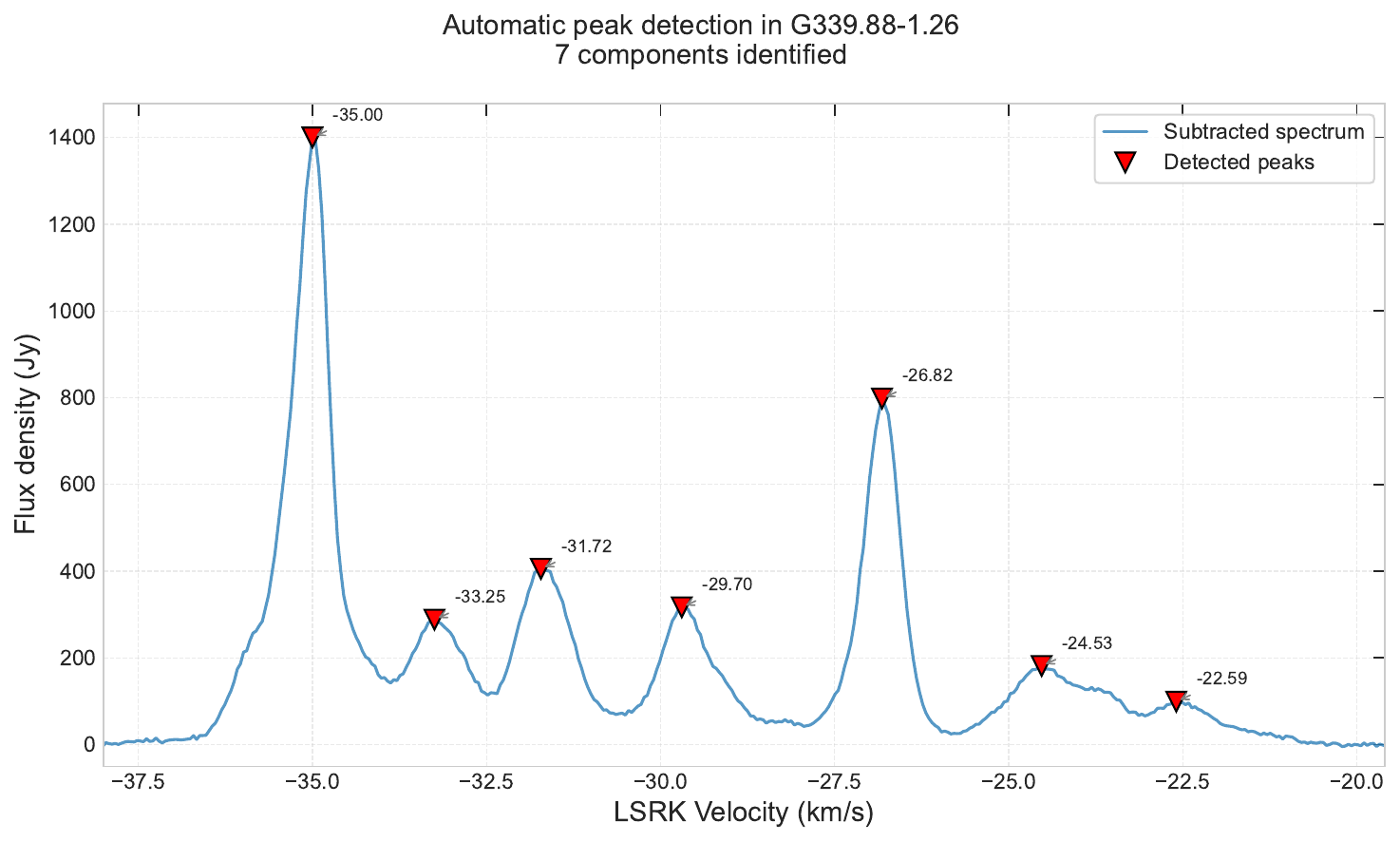}
\caption{Calibrated 6.7~GHz methanol maser spectrum of G339.884$-$1.259 from GRAO. The flux density (Jy) versus LSRK velocity (km~s$^{-1}$) is shown after ON-OFF subtraction and velocity alignment. Red diamonds indicate seven automatically detected peaks. The strongest peak reaches 1400~Jy at $-35.0$~km~s$^{-1}$, with weaker components between $-33$ and $-23$~km~s$^{-1}$. These peaks provide initial guesses for the Bayesian MCMC decomposition.}
\label{fig:fig_peak_detection}
\end{figure*}

We use the \texttt{emcee} software package \citep{2013PASP..125..306F}, which implements an affine‑invariant ensemble sampler. This variant uses multiple walkers and updates each walker using the current positions of other walkers, making it more efficient for correlated parameter spaces. The algorithm is fully described in \cite{2010CAMCS...5...65G}.
To assess whether the MCMC chains have converged to the stationary distribution, we compute the Gelman-Rubin \(\hat{R}\) statistic \citep{1992StaSc...7..457G}. For each parameter we run multiple chains (here, 100 walkers are treated as independent chains after burn‑in). Let \(W\) be the within‑chain variance and \(B\) the between‑chain variance; then
\begin{equation}
    \hat{R} = \sqrt{\frac{n-1}{n} + \frac{B}{nW}},
\end{equation}
where \(n\) is the number of samples per chain (after burn‑in and thinning). Values of \(\hat{R} < 1.05\) indicate convergence. All parameters in our fits satisfy this criterion. Additionally, we visually inspect the trace plots (see Section~\ref{sec:results}) to ensure no drift or poor mixing.
From the thinned posterior samples we report the median as the point estimate and the 16th and 84th percentiles as the 68 per cent credible interval (analogous to \(\pm 1\sigma\) for a normal distribution).
The number of spectral components is unknown. We automatically detect peaks using the \texttt{scipy.signal.find\_peaks} function with the following parameters: height = \(0.05\max(y)\), prominence = \(0.02\max(y)\), width = 2 samples, and a minimum distance of 5 samples. This procedure identifies seven distinct peaks (see Fig.~\ref{fig:fig_peak_detection}). For each peak we set the initial amplitude to the peak height, the initial centre to the peak velocity, and the initial width parameters (\(\sigma\) for Gaussian, \(\gamma\) for Lorentzian, both for Voigt) from the peak width at half maximum, converted via the appropriate FWHM formula.

A preliminary least‑squares fit using the \texttt{lmfit} library \citep{newville2025lmfit} provides a maximum‑likelihood estimate that serves as the starting point for the MCMC walkers.

\subsection{Model comparison}

To decide which functional family (Gaussian, Lorentzian, or Voigt) best describes the data, we compare several goodness‑of‑fit statistics. Each statistic highlights different aspects of the fit quality.

The coefficient of determination, \(R^2\), measures the proportion of the variance in the data that is explained by the model:
\begin{equation}
    R^2 = 1 - \frac{\sum_i (y_i - f_M(v_i))^2}{\sum_i (y_i - \bar{y})^2},
\end{equation}
where \(\bar{y}\) is the mean of the observed flux densities. Values close to 1 indicate a nearly perfect reproduction of the data. However, \(R^2\) alone does not penalise model complexity.

The chi‑square statistic, \(\chi^2 = \sum_i (y_i - f_M(v_i))^2 / \varepsilon^2\), and the reduced chi‑square, \(\chi^2_{\nu} = \chi^2 / (N - k)\), provide absolute measures of the discrepancy between model and data, scaled by the noise variance \(\varepsilon^2\) and the degrees of freedom \(\nu = N - k\). A reduced chi‑square near 1 suggests that the model fits the data within the estimated uncertainties; values substantially larger than 1 indicate under‑fitting or underestimated noise.

For a more rigorous comparison that penalises extra parameters, we use the Akaike Information Criterion (AIC) and the Bayesian Information Criterion (BIC). For a model with \(k\) free parameters and maximum likelihood \(\hat{L}\) (evaluated from the least‑squares fit assuming Gaussian errors),
\begin{align}
    \mathrm{AIC} &= 2k - 2\ln\hat{L}, \label{eq:aic} \\
    \mathrm{BIC} &= k\ln N - 2\ln\hat{L}. \label{eq:bic}
\end{align}
Both criteria balance goodness of fit against model complexity; lower values indicate a more parsimonious model. BIC imposes a stronger penalty for additional parameters when the number of data points \(N\) is large.
The number of free parameters per component differs among the three models (see Table~\ref{tab:params_count}). The total number of parameters is \(k = K \times (\text{parameters per component}) + 1\), where the extra term accounts for the noise variance \(\varepsilon^2\), which is also estimated from the data. Here \(K = 7\) is the number of detected components.
We compute \(R^2\), \(\chi^2\), \(\chi^2_{\nu}\), AIC, and BIC for each model (Gaussian, Lorentzian, Voigt) from the least‑squares fits. The model with the lowest AIC and BIC, a reduced chi‑square closest to 1, and the highest \(R^2\) is considered statistically preferred. 

\begin{table}
\centering
\caption{Number of free parameters per component for each model family.}
\label{tab:params_count}
\begin{tabular}{lcc}
\hline
Model & Parameters per component & Symbol \\
\hline
Gaussian & 3 & \(A, \mu, \sigma\) \\
Lorentzian & 3 & \(A, \mu, \gamma\) \\
Voigt & 4 & \(A, \mu, \sigma, \gamma\) \\
\hline
\end{tabular}
\end{table}


\section{RESULTS} \label{sec:results}

The calibrated spectrum of the 6.7\,GHz methanol maser in G339.884$-$1.259 reveals a complex, multi-component structure spanning LSRK velocities from approximately $-35$ to $-23$\,km\,s$^{-1}$ (Fig.~\ref{fig:fig_peak_detection}). Seven statistically significant peaks are identified, consistent with the expected fragmentation of maser emission into discrete velocity-coherent regions within high-mass star-forming environments. The strongest feature, located near $-35$\,km\,s$^{-1}$, reaches a peak flux density exceeding 1300\,Jy, while secondary components at higher velocities generally exhibit lower intensities, although the second-brightest feature occurs at an intermediate velocity.
This distribution reflects a structured velocity field rather than a continuous emission profile, suggesting that the maser emission traces physically distinct parcels of gas rather than a single homogeneous region.

\begin{figure*}
\centering
\includegraphics[width=\textwidth]{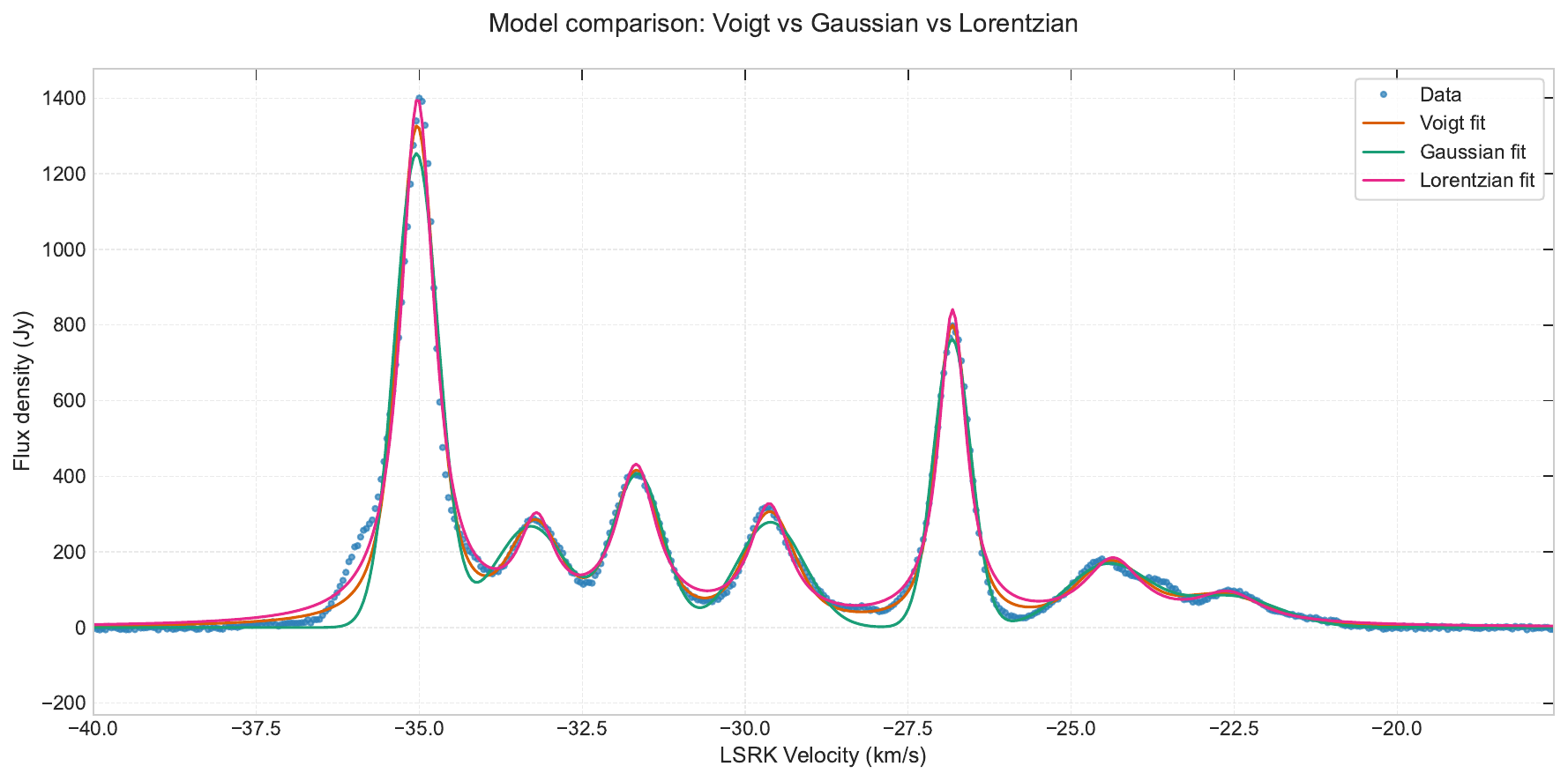}
\caption{Comparison of Gaussian, Lorentzian, and Voigt composite fits overlaid on the observed spectrum, illustrating differences in core and wing representation.
}
\label{fig:fig_model_comparison_fit}
\end{figure*}

A direct comparison between Gaussian, Lorentzian, and Voigt representations demonstrates systematic differences in how each model captures the observed line shapes (Fig.~\ref{fig:fig_model_comparison_fit}). The Gaussian model reproduces the core of each feature reasonably well but underestimates flux in the line wings. The Lorentzian model captures broader wings but introduces excess flux in regions where the data exhibit sharper cut-offs. The Voigt profile provides a consistent compromise, reproducing both the narrow cores and extended wings without introducing systematic deviations across the spectrum. This qualitative behaviour is supported quantitatively by the model comparison statistics in Table~\ref{tab:model_comparison}. The Voigt model achieves the lowest AIC (19767.5487) and BIC (19900.2218), alongside the smallest RMSE (11.1106\,Jy) and the highest coefficient of determination ($R^{2}=0.9854$). The Gaussian model shows a significantly larger information criterion penalty, while the Lorentzian model yields intermediate performance but remains statistically disfavoured. These results indicate that the additional degrees of freedom in the Voigt model are justified by the data and are not merely compensating for noise.

\begin{figure*}
\centering
\includegraphics[width=\textwidth]{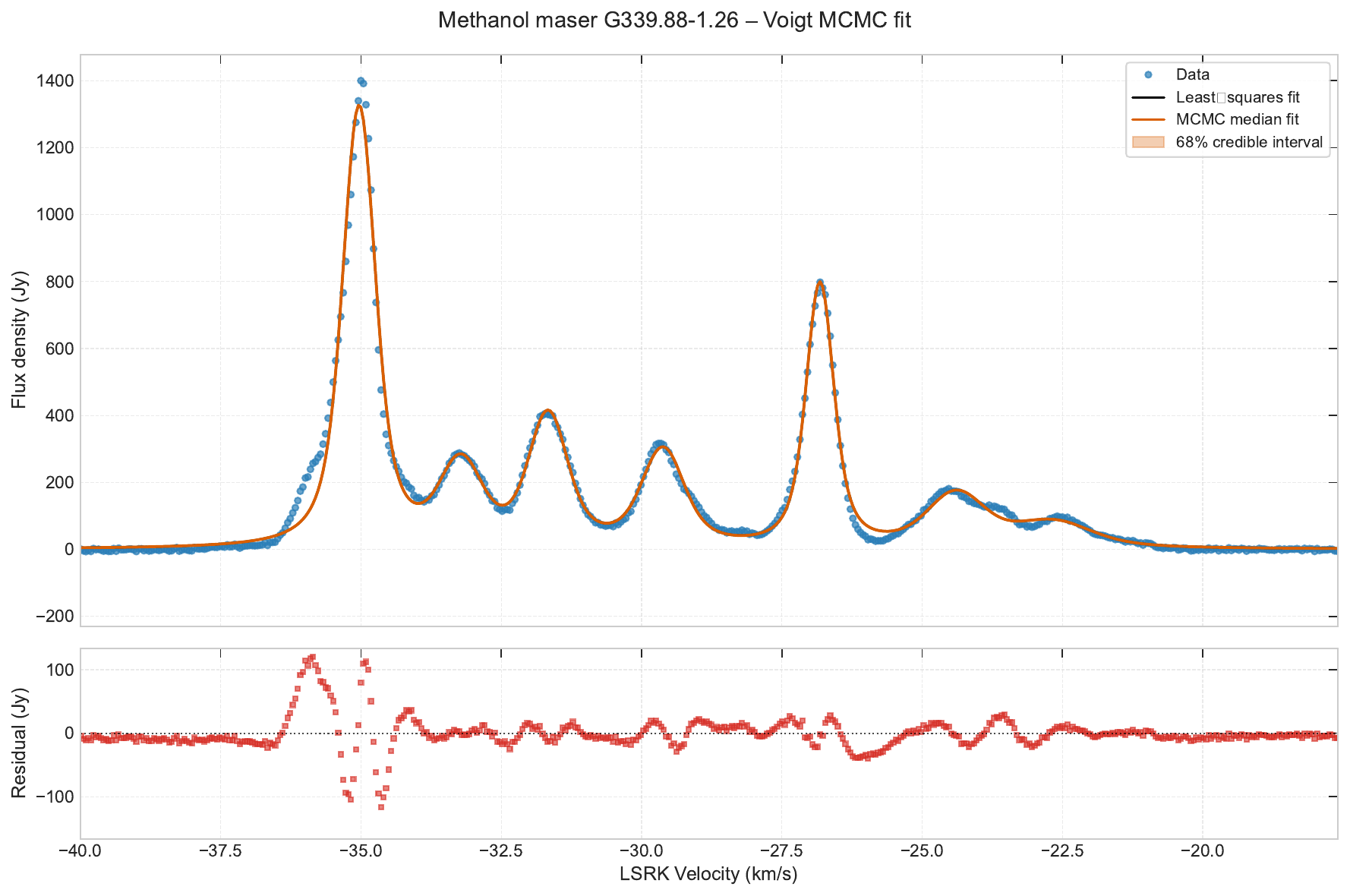}
\caption{Best-fitting Voigt model with MCMC median solution and 68\,per\,cent credible interval; residuals shown in the lower panel. The lower panel is displayed on an expanded vertical scale to highlight residual spectral structure.
}
\label{fig:best_fit_voigt}
\end{figure*}

\begin{table*}
\centering
\caption{Statistical comparison of Gaussian, Lorentzian, and Voigt models based on goodness-of-fit and information criteria.}
\label{tab:model_comparison}
\begin{tabular}{lcccccc}
\hline
Model & RMSE (Jy) & $R^{2}$ & $\chi^{2}$ & $\chi^{2}_{\nu}$ & AIC & BIC \\
\hline
Voigt      & 11.1106 & 0.9854 & 505636.8283 & 124.0827 & 19767.5487 & 19900.2218 \\
Gaussian   & 15.1783 & 0.9727 & 943644.8046 & 231.5693 & 22323.1707 & 22455.8438 \\
Lorentzian & 11.3439 & 0.9847 & 527089.4421 & 129.3471 & 19937.7440 & 20070.4171 \\
\hline
\end{tabular}
\end{table*}

The detailed Voigt fit (Fig.~\ref{fig:best_fit_voigt}) confirms that the Bayesian MCMC solution closely follows the least-squares estimate while providing a principled uncertainty envelope. The 68\,per\,cent credible interval remains narrow across most velocity channels, indicating that the data strongly constrain the model parameters. Residuals are centred around zero with no large-scale systematic trends, although localised deviations are present near the strongest component. These residual structures are small compared to the peak flux and are more likely associated with unresolved spectral blending, partial overlap of masing clouds along the line of sight, internal velocity gradients, or departures from the assumed analytic profile shapes. The reduced $\chi^{2}$ values exceeding unity for all models suggest that either the noise is slightly underestimated or that intrinsic spectral complexity is not fully captured by a finite number of components.

Posterior parameter estimates (Table~\ref{tab:voigt_parameters}) show that the maser components are narrowly distributed in velocity space with centre uncertainties typically below $0.01$\,km\,s$^{-1}$. The smallest uncertainties occur for the brightest components, reflecting the strong signal-to-noise ratio. Full width at half-maximum values range from approximately $0.59$ to $1.68$\,km\,s$^{-1}$, consistent with previously reported linewidths for Class~II methanol masers \citep{2011ApJ...730...55P,2010A&A...517A..56F}. For comparison, the thermal linewidth expected for methanol under typical conditions in high-mass star-forming regions is narrower than the broadest observed components. Assuming kinetic temperatures in the range $\sim 30$--$70$\,K reported for G339.884$-$1.259 by \citet{2019ApJ...873...73Z}, the corresponding thermal FWHM for CH$_3$OH is approximately $\sim 0.27$--$0.41$\,km\,s$^{-1}$, increasing to $\sim 0.53$\,km\,s$^{-1}$ at 200\,K. The observed linewidths therefore likely reflect a combination of thermal motions, velocity coherence, turbulence, saturation effects, radiative transfer processes, unresolved blending, and internal velocity gradients rather than purely thermal broadening alone. Previous polarimetric VLBI observations of G339.884$-$1.259 suggest that the brightest feature near $-35$\,km\,s$^{-1}$ may be partially saturated \citep{2008A&A...480..767D}. Overall, the linewidth distribution supports a scenario in which maser amplification occurs along structured, velocity-coherent paths within a dynamically complex medium.

\begin{table*}
\centering
\caption{Posterior parameter estimates for individual maser components derived from the Voigt MCMC model.}
\label{tab:voigt_parameters}
\begin{tabular}{ccccccc}
\hline
Component & Peak Flux (Jy) & Flux err (Jy) & Centre (km s$^{-1}$) & Centre err (km s$^{-1}$) & FWHM (km s$^{-1}$) & FWHM err (km s$^{-1}$) \\
\hline
1 & 75.4760 & 2.8495 & -22.6569 & 0.0462 & 1.6783 & 0.1219 \\
2 & 159.4206 & 3.4850 & -24.3925 & 0.0174 & 1.2595 & 0.0482 \\
3 & 786.2249 & 4.6005 & -26.8204 & 0.0017 & 0.5949 & 0.0046 \\
4 & 288.2587 & 3.6959 & -29.6182 & 0.0059 & 0.9232 & 0.0159 \\
5 & 389.3503 & 3.8633 & -31.6640 & 0.0043 & 0.8476 & 0.0124 \\
6 & 243.7701 & 3.6964 & -33.2206 & 0.0073 & 0.9289 & 0.0210 \\
7 & 1315.4683 & 4.2212 & -35.0321 & 0.0011 & 0.7070 & 0.0031 \\
\hline
\end{tabular}
\end{table*}

The spatially unresolved but spectrally distinct components can be interpreted within the established framework of methanol maser excitation in massive star-forming regions. The maser components exhibit a structured velocity distribution spanning approximately 13\,km\,s$^{-1}$, consistent with an organised kinematic environment such as rotation, outflow, or overlapping velocity-coherent structures. Earlier interferometric studies of G339.884$-$1.259 reported linear maser distributions and velocity gradients that were initially attributed to a rotating disc but later reconsidered in favour of a more complex geometry involving outflows or multiple embedded sources \citep{2000ApJS..130..437D,1996PhDT........15E}. Polarimetric VLBI observations by \citet{2008A&A...480..767D} similarly indicate that the source geometry is unlikely to be explained by a simple rotating disc alone. The present spectral decomposition supports this revised interpretation, with the presence of multiple narrow components and small velocity separations favouring maser amplification within a structured and dynamically non-uniform environment rather than a single dynamically simple system.

\begin{figure*}
\centering
\includegraphics[width=0.8\textwidth]{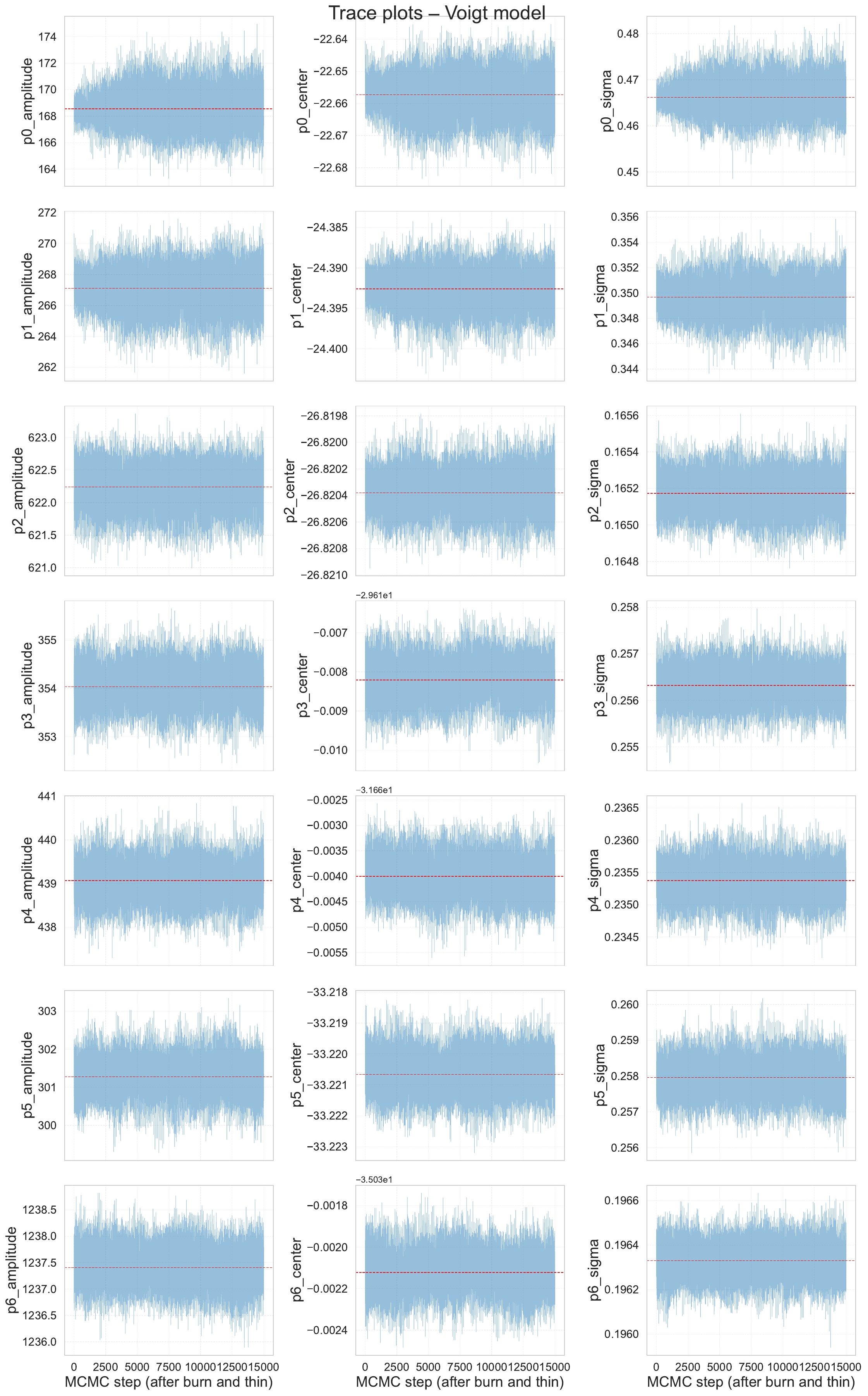}
\caption{Representative MCMC trace plots for selected Voigt parameters (amplitude, centre, and width) illustrating convergence and stable mixing.}
\label{fig:trace_plots_voigt}
\end{figure*}

The MCMC trace plots (Fig.~\ref{fig:trace_plots_voigt}) exhibit stationary behaviour with no long-term drift and rapid fluctuations about stable means, consistent with convergence of the sampler. This visual assessment is supported by the Gelman--Rubin statistic reported in Section~3, with all parameters satisfying $\hat{R} < 1.05$, indicating that between-chain and within-chain variances are comparable and that the posterior has been adequately explored. 

\begin{figure*}
\centering
\includegraphics[width=\textwidth]{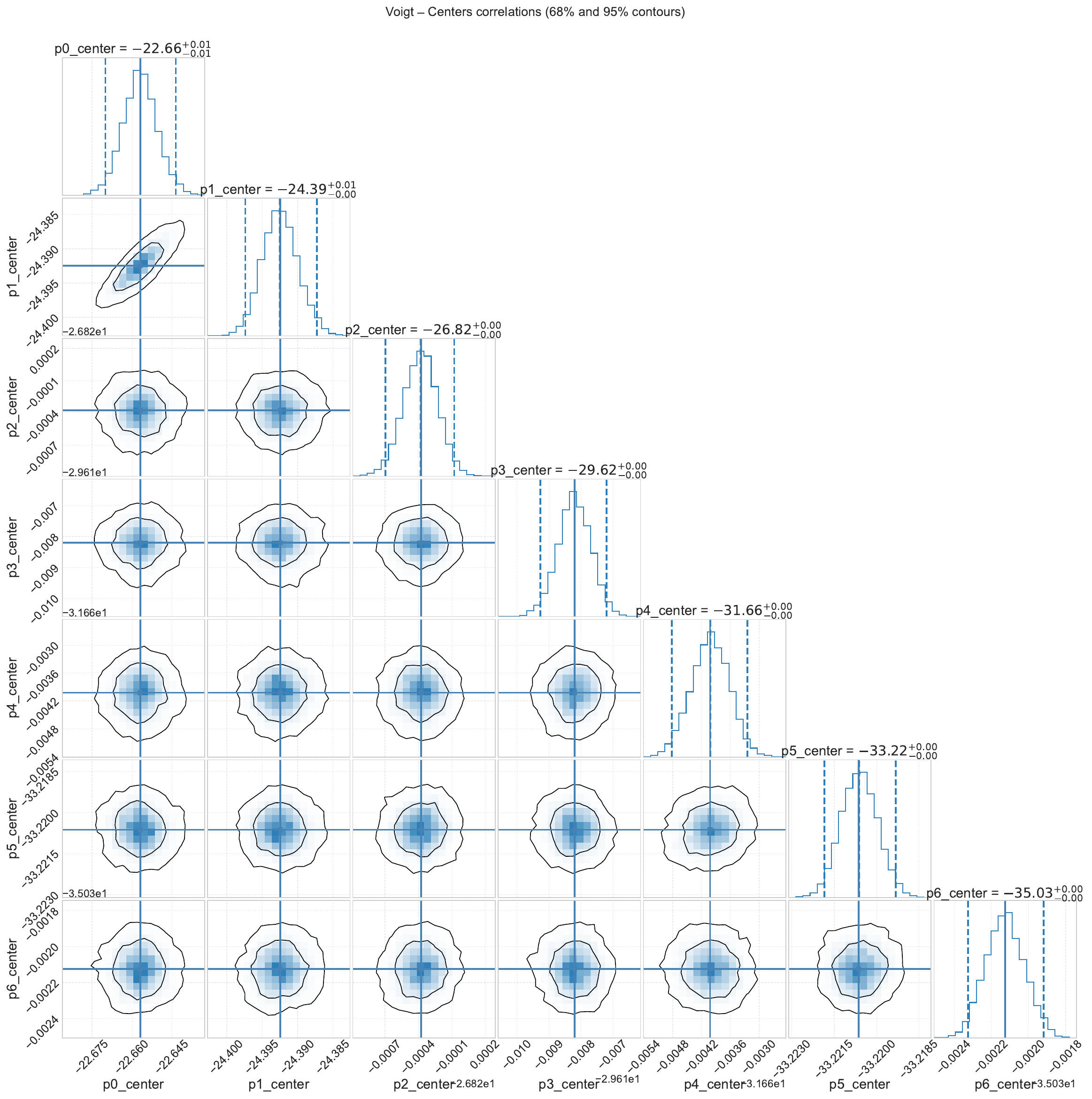}
\caption{Corner plot of the posterior distributions for the central velocities of the Voigt components. The diagonal panels show the marginalised one-dimensional distributions with median values and credible intervals, while the off-diagonal panels display two-dimensional projections that quantify correlations between component centres. The narrow, well-separated distributions indicate that the velocities are tightly constrained and largely independent, supporting the identification of distinct kinematic components within the maser spectrum.}
\label{fig:corner_voigt_centers}
\end{figure*}

The corner plots (Fig.~\ref{fig:corner_voigt_centers}) provide complementary information on parameter constraints and covariance structure. The marginal posterior distributions for the component centres are narrow relative to the spectral resolution and are well separated in velocity space, indicating that the kinematic components are individually resolved. Off-diagonal panels show minimal elongation, suggesting that degeneracy between neighbouring centres is limited. 

In contrast, weak correlations are visible between the amplitudes of adjacent components, which is expected in regions of partial spectral blending where flux can be redistributed between nearby features. The width parameters exhibit only modest covariance, implying that the data constrain line broadening sufficiently to distinguish between Gaussian and Lorentzian contributions within the Voigt formalism. Taken together, these diagnostics indicate that the parameter estimates are both statistically stable and physically interpretable within the adopted model.

From a phenomenological perspective, the preference for the Voigt profile indicates that the observed maser features are not perfectly represented by purely Gaussian functions. The additional wing structure captured by the Lorentzian term may reflect a combination of unresolved spectral blending, velocity gradients, saturation effects, radiative transfer processes, turbulence, or other departures from ideal Gaussian behaviour. The present data do not uniquely isolate the physical origin of these non-Gaussian contributions, particularly in the absence of spatially resolved interferometric constraints. The statistical preference for the Voigt representation therefore, indicates that a more flexible profile model is required to reproduce the observed spectral morphology, rather than uniquely demonstrating classical damping broadening within the masing medium.

\section{Discussion}\label{sec:discuss}

The spectral decomposition presented in this work provides a physically informative framework for interpreting the methanol maser source G339.884$-$1.259 beyond a purely descriptive characterisation of the observed spectrum. The presence of multiple narrow velocity components is consistent with the established picture of Class~II methanol masers arising in compact, velocity-coherent regions embedded within high-mass star-forming environments. Such components are generally interpreted as elongated amplification paths in which sufficiently small velocity gradients permit coherent maser gain. The separation of these components in velocity space therefore suggests that the masing medium is structured rather than continuous, with multiple physically distinct regions contributing independently to the observed emission. This interpretation is consistent with interferometric studies of methanol masers, where individual maser spots are often spatially clustered yet kinematically distinct, tracing organised motions associated with discs, shocks, or outflows.

The statistical preference for the Voigt profile over purely Gaussian or Lorentzian representations indicates that the observed spectra contain measurable departures from ideal Gaussian structure. However, the Lorentzian contribution should not be interpreted uniquely as evidence for collisional or radiative damping within the maser medium. In methanol masers, non-Gaussian wings may arise from several effects, including unresolved velocity gradients, partial saturation, radiative transfer effects, spectral blending of nearby maser spots, turbulence, temperature and density gradients affecting the inversion conditions, or damping-related processes. Theoretical studies predict that unsaturated methanol masers tend towards approximately Gaussian line profiles, while saturation and radiative transfer effects can produce departures from ideal Gaussian behaviour \citep{Gray2012,2005MNRAS.360..533C,2002MNRAS.331..521C,elitzur1992astronomical}. High-angular-resolution studies further demonstrate that methanol masers frequently exhibit complex small-scale velocity structure and internal gradients that cannot be fully captured by spatially unresolved single-dish spectra \citep{2024IAUS..380..172M}. The present analysis therefore treats the Voigt profile primarily as a flexible phenomenological representation of the observed spectral morphology rather than as a direct physical decomposition of isolated microscopic broadening mechanisms.

Interpretation of the velocity distribution nevertheless requires caution, particularly given the absence of spatial resolution in single-dish observations. The maser components exhibit a structured velocity distribution spanning approximately 13\,km\,s$^{-1}$, which may plausibly arise from organised kinematic environments such as rotation, outflow, shocks, or overlapping velocity-coherent structures. Earlier high-resolution studies of G339.884$-$1.259 initially favoured a rotating disc interpretation but later identified multiple infrared sources and more complex morphology, indicating that a single dynamical model is insufficient \citep{2000ApJS..130..437D,1996PhDT........15E}. Polarimetric VLBI observations by \citet{2008A&A...480..767D} similarly suggest that the source geometry cannot be explained adequately by a simple rotating disc alone. The present decomposition supports this revised interpretation by demonstrating that the observed spectral profile is better represented by multiple partially overlapping velocity components rather than by a single smooth kinematic structure. While the observed velocity ordering is suggestive of organised motion, it does not uniquely distinguish between competing dynamical scenarios in the absence of spatial information.

The linewidths inferred from the Voigt decomposition provide additional constraints on the physical state of the masing gas. The observed linewidths are comparable to or moderately larger than the expected thermal linewidths for methanol under typical conditions in high-mass star-forming regions and therefore do not require strong pressure broadening. Unsaturated maser amplification in an entirely quiescent medium would likely produce narrower profiles than those observed here, whereas turbulence substantially exceeding $\sim1$\,km\,s$^{-1}$ would tend to disrupt coherent amplification. The inferred linewidths therefore likely reflect a combination of thermal motions, velocity coherence, turbulence, unresolved blending, internal velocity gradients, partial saturation, and radiative transfer effects within the masing environment. Previous polarimetric VLBI observations suggest that the brightest feature near $-35$\,km\,s$^{-1}$ may be partially saturated \citep{2008A&A...480..767D}, while weaker components are likely influenced more strongly by blending and local velocity structure. Higher angular resolution observations will be required to determine which of these effects dominates the observed wing structure and component asymmetries.

Although the statistical diagnostics favour the adopted model, several limitations must be considered when interpreting the results. The reduced $\chi^{2}$ values reported for all models are significantly larger than unity, indicating that the residual variance exceeds that expected from the assumed noise level. Consequently, although the Voigt model is statistically preferred relative to the Gaussian and Lorentzian alternatives, none of the tested models should be regarded as a complete physical description of the maser emission. This discrepancy may partly arise from uncertainties in the instrumental noise estimate, but it may also reflect intrinsic spectral complexity that cannot be fully captured by analytic symmetric profile models. In particular, unresolved substructure within individual maser features, partial overlap of masing clouds along the line of sight, or internal velocity gradients may contribute to the residual patterns observed near the strongest components. These residual structures suggest that higher spectral resolution and multi-epoch observations would be valuable for disentangling the detailed temporal and kinematic behaviour of the source.

The Bayesian framework adopted in this study provides a transparent and statistically rigorous way to quantify these uncertainties and assess parameter identifiability. The posterior distributions demonstrate that the central velocities are tightly constrained and largely separable, while correlations between amplitudes remain mostly confined to neighbouring blended components. This behaviour indicates that the decomposition is not dominated by severe parameter degeneracy, although some degree of coupling is unavoidable in strongly blended regions. The ability of the Voigt formalism to reproduce residual wing structure more effectively than purely Gaussian or Lorentzian models further suggests that the spectra contain measurable information beyond simple symmetric Doppler broadening.

Methodologically, the results highlight the advantages of Bayesian MCMC approaches for maser spectroscopy. Traditional least-squares fitting methods generally provide a single optimum solution but do not fully capture parameter uncertainties, posterior correlations, or model ambiguity. In contrast, the probabilistic framework adopted here allows direct assessment of parameter precision, uncertainty propagation, and model robustness. This is particularly important for complex blended spectra where subjective choices in component selection can influence the inferred physical interpretation. Embedding the decomposition within a Bayesian framework therefore improves both the reproducibility and interpretability of the analysis.

The broader implication is that statistically rigorous spectral modelling can substantially enhance the interpretation of single-dish maser observations, even in the absence of spatial resolution. At the same time, the present results emphasise that spectral decomposition alone cannot uniquely determine the three-dimensional geometry or dynamical state of the source. Future high-angular-resolution and multi-epoch observations will therefore be essential for disentangling the spatial structure, temporal variability, and kinematic evolution of individual maser components \citep{2025A&A...701A..28W,2023A&A...671A.135K}. Complementary observations of other maser species may further constrain the physical conditions and dynamical environment within the star-forming region. In this context, Bayesian inference frameworks offer a promising pathway for integrating spectroscopic, temporal, and interferometric information within a unified statistically consistent analysis.

\section{Conclusion} \label{sec:conc}

This study has presented a Bayesian framework for the decomposition of the 6.7\,GHz methanol maser spectrum of G339.884$-$1.259 using composite line profile models. By embedding the spectral fitting within a MCMC scheme, the analysis moves beyond conventional least-squares approaches and provides a statistically coherent characterisation of both parameter estimates and their associated uncertainties. The resulting posterior distributions enable direct assessment of parameter precision and identifiability, which is essential for interpreting complex, blended maser spectra.

The comparative analysis of Gaussian, Lorentzian, and Voigt representations demonstrates that the Voigt formalism provides a statistically preferred description of the observed maser spectra relative to purely Gaussian or Lorentzian models. This outcome indicates that the observed maser spectra exhibit measurable departures from purely Gaussian structure and that the Voigt formalism provides a more flexible statistical representation of the line morphology. The results therefore, reinforce the importance of adopting flexible and physically informed profile models when analysing high signal-to-noise maser spectra.

Despite the overall consistency of the model with the observed data, the analysis highlights limitations inherent to single-dish spectral decomposition. In particular, the presence of structured residuals and elevated goodness-of-fit statistics indicates that the adopted models do not fully capture all aspects of the emission, likely due to unresolved substructure or departures from the assumed noise properties. These factors underscore the need for caution when interpreting fine-scale spectral features and emphasise that statistical adequacy does not necessarily imply complete physical description.

The Bayesian methodology employed here provides a general and reproducible framework that can be extended to other maser sources and molecular line observations. Its strength lies in the ability to quantify uncertainty and parameter correlations in a principled manner, thereby reducing the subjectivity associated with traditional fitting techniques. Future work would benefit from integrating this approach with high-resolution interferometric data, enabling direct association between spectral components and spatially resolved maser structures. Such a combination would offer a more complete understanding of the kinematics and physical conditions in massive star-forming regions.

\section*{Acknowledgement}

The authors acknowledge that the data used in this work were obtained during the Development in Africa with Radio Astronomy (DARA) training programme. We are grateful to the DARA initiative, a joint UK–South African development project, for its role in building capacity and advancing radio astronomy research across the African continent.
This work was carried out during the Phase II engineering stage of the Ghana Radio Astronomy Observatory (GRAO). We thank the technical engineers from the South African Radio Astronomy Observatory (SARAO) for their continued support in the development and operation of the telescope. Their expertise has been instrumental in advancing the facility towards a modern, high-performance instrument.
We acknowledge Prof. Melvin Hoare for his sustained support of the DARA programme and its contribution to the growth of radio astronomy in Africa. We also recognise the Regents of the University of California, on behalf of the University of California, Santa Cruz (UCSC), for their collaboration in advancing joint research and educational activities in astronomy and astrophysics, with particular emphasis on the development and scientific utilisation of the Ghana Radio Astronomy Observatory.

\section*{Data Availability}

The data underlying this study were obtained using the Ghana Radio Astronomy Observatory (GRAO) during the Development in Africa with Radio Astronomy (DARA) training programme. The processed data products and analysis scripts used in this work are available from the corresponding author upon reasonable request. 
Raw observational data may be subject to institutional access policies during the facility's ongoing engineering phase but can be made available where permitted.



\bibliographystyle{mnras}
\bibliography{mnras_ref} 


\appendix

\section{Supplementary Posterior Distributions for Voigt Model Parameters}

The posterior distributions presented here complement the convergence diagnostics shown in the main text. While trace plots demonstrate that the MCMC chains have reached stationarity, the corner plots provide insight into parameter identifiability and correlations. In particular, they allow assessment of whether individual spectral components are uniquely constrained or affected by degeneracies arising from line blending.

\begin{figure*}
\centering
\includegraphics[width=\textwidth]{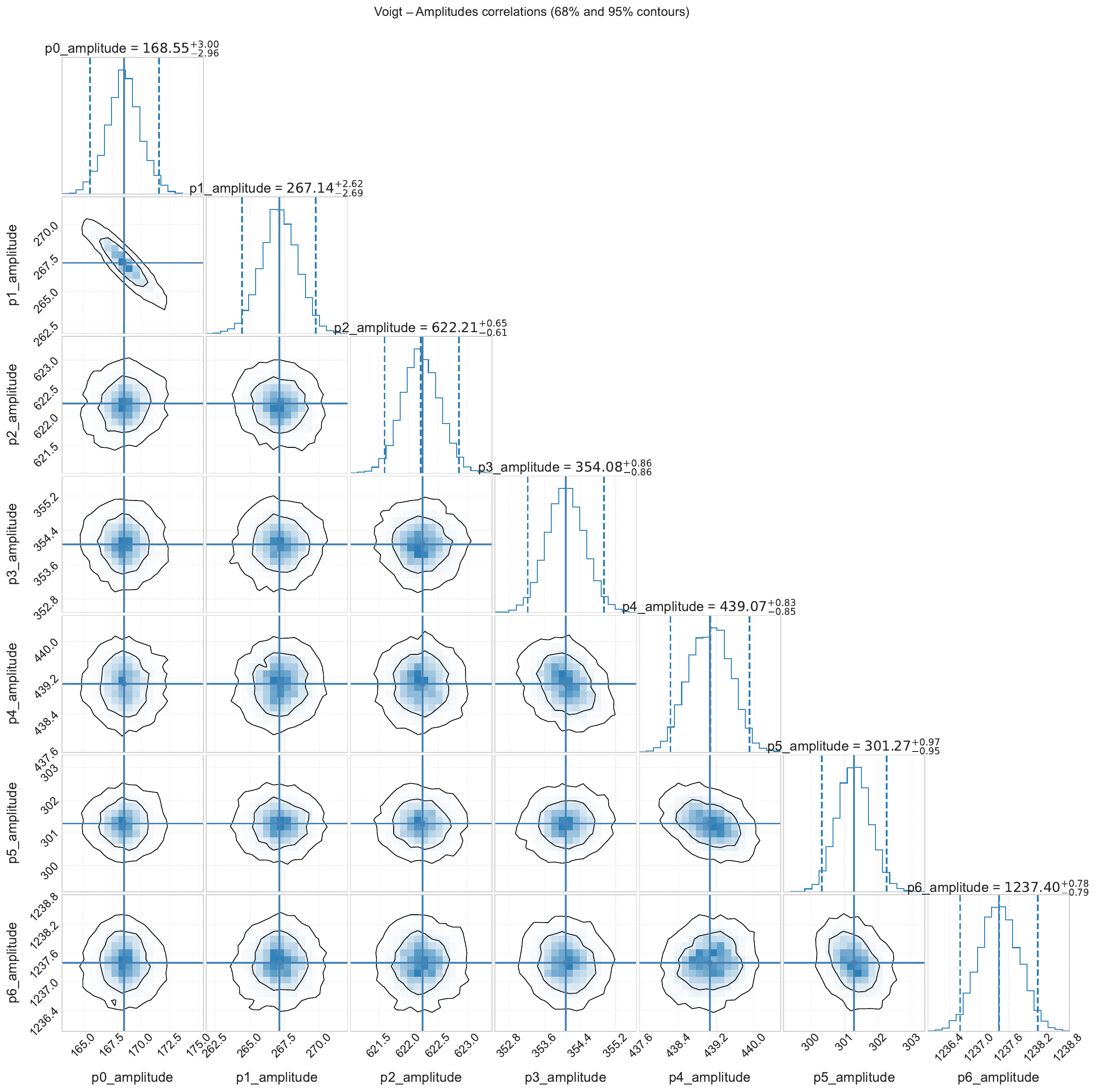}
\caption{Corner plot of posterior distributions for the Voigt component amplitudes. Diagonal panels show the marginalised one-dimensional distributions with median values and credible intervals, while off-diagonal panels display two-dimensional projections highlighting correlations between amplitudes of different components. The generally weak covariance indicates that individual component fluxes are well constrained despite partial spectral blending.}
\label{fig:corner_voigt_amplitudes}
\end{figure*}

\begin{figure*}
\centering
\includegraphics[width=\textwidth]{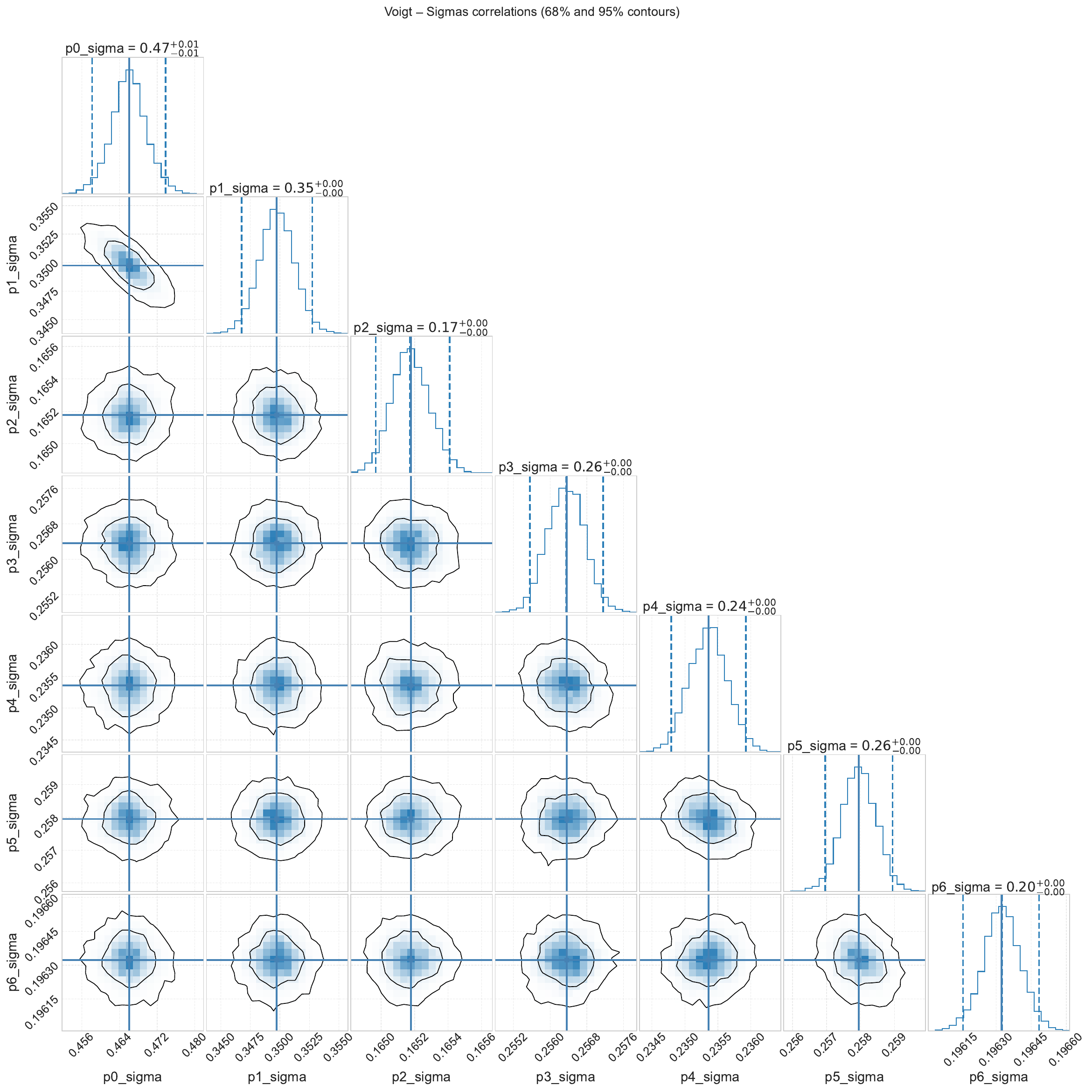}
\caption{Corner plot of posterior distributions for the Gaussian width parameters ($\sigma$) of the Voigt components. The narrow, symmetric marginal distributions indicate strong constraints on linewidths, while the limited off-diagonal structure suggests weak correlations between components. These results support the stability of the decomposition and the ability of the data to resolve individual velocity-coherent features.}
\label{fig:corner_voigt_sigmas}
\end{figure*}

\bsp	
\label{lastpage}
\end{document}